%% file: main.tex
\documentclass[conference]{IEEEtran}
\usepackage[numbers,sort&compress,square]{natbib}

\usepackage{amsmath} 
\usepackage{mathtools}

\usepackage[text-rm]{siunitx} 
\usepackage{booktabs} 
\usepackage{multirow} 
\usepackage{arydshln} 
\usepackage[table]{xcolor} 
\usepackage{tabularx} 
\usepackage{xspace} 
\usepackage{stmaryrd}
\usepackage[hyphens]{url}
\usepackage{wrapfig}

\usepackage[scaled]{helvet} 
\newcommand{\hel}{\fontfamily{phv}\selectfont}

\newcommand{\ignore}[1]{}

\newcommand{\jk}[1]{{\color{black}#1}}
\newcommand{\jktwo}[1]{{\color{black}#1}}
\newcommand{\jkthree}[1]{{\color{black}#1}}
\newcommand{\jkfour}[1]{{\color{black}#1}}

\newcommand{\module}[3]{{{\hel #1}$_{\mathrm{#2}}^{\mathrm{#3}}$}\xspace}

\ifCLASSINFOpdf
   \usepackage[pdftex]{graphicx}
\else
\fi

\usepackage{algorithm}
\usepackage{algorithmicx} 
\floatname{algorithm}{Template}

\usepackage{caption}
\usepackage{subcaption}
\begin{document}
\bstctlcite{IEEEexample:BSTcontrol} 
%
\title{RowHammer: A Retrospective}


\newcommand{\affilCMU}[0]{\textsuperscript{$\ddagger$}}
\newcommand{\affilETH}[0]{\textsuperscript{\S}}

\author{
{Onur Mutlu\affilETH\affilCMU}\qquad%
{Jeremie S. Kim\affilCMU\affilETH}\qquad\\%
{\affilETH ETH Z{\"u}rich \qquad \affilCMU Carnegie Mellon University }%
\vspace{-5pt}%
}


%


\renewcommand\IEEEkeywordsname{Index Terms}

\maketitle

\thispagestyle{plain}
\pagestyle{plain}

\input{abstract}


\IEEEpeerreviewmaketitle
\input{introduction}
\input{rowhammer_summary}

\input{related}

\input{future}

\input{conclusion}


\section*{Acknowledgments}

This paper is based on two previous papers we have written on RowHammer, one that first scientifically introduced and analyzed the phenomenon in ISCA 2014~\cite{rowhammer-isca2014} and the other that
provided an analysis and future outlook on RowHammer~\cite{onur-date17}. \jkthree{The presented work} is a result
of the research done together with many students and collaborators over the
course of the past eight years. In particular, three PhD theses have shaped the
understanding that led to this work. These are Yoongu Kim's thesis entitled
"Architectural Techniques to Enhance DRAM Scaling"~\cite{yoongu-thesis}, Yu Cai's thesis entitled "NAND Flash Memory: Characterization, Analysis, Modeling and Mechanisms"~\cite{yucai-thesis} and his continued follow-on work after his thesis, summarized in~\cite{cai2017error, cai2017errors}, and Donghyuk Lee's thesis entitled "Reducing DRAM Latency at Low Cost
by Exploiting Heterogeneity"~\cite{donghyuk-thesis-arxiv16}. We also acknowledge various funding agencies (NSF, SRC, ISTC, CyLab) and industrial partners (AliBaba, AMD, Google,
Facebook, HP Labs, Huawei, IBM, Intel, Microsoft, Nvidia, Oracle, Qualcomm,
Rambus, Samsung, Seagate, VMware) who have supported the presented and
other related work in \jkthree{our} group generously over the years. 

The first version of
\jkthree{the talk associated with this paper} was delivered at a CMU CyLab Partners Conference in September
2015. \jkthree{Other versions of the talk were} delivered as part of an Invited Session
at DAC 2016, with a collaborative accompanying paper entitled ''Who Is the
Major Threat to Tomorrow\jkfour{'}s Security? You, the Hardware Designer''~\cite{dac-invited-paper16}, \jkthree{at DATE 2017~\cite{onur-date17},} and at the Top Picks in Hardware and
Embedded Security workshop, co-located with ICCAD 2018~\cite{TopPicks}, where RowHammer was selected as a Top Pick among hardware and embedded security papers
published between 2012-2017. \jkthree{The most recent version of the associated talk was delivered at COSADE 2019~\cite{mutlu2019rowhammer}}.

\Urlmuskip=0mu plus 1mu\relax

{
  \scriptsize 
  \renewcommand{\baselinestretch}{0.95}
  \let\OLDthebibliography\thebibliography
  \renewcommand\thebibliography[1]{
    \OLDthebibliography{#1}
    \setlength{\parskip}{0pt}
    \setlength{\itemsep}{0pt}
  }
  \bibliographystyle{IEEEtranS}
  \bibliography{paper} 
}

 \begin{wrapfigure}{l}{25mm} 
    \includegraphics[width=1in,height=1.25in,clip,keepaspectratio]{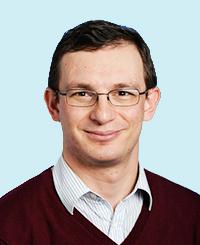}
  \end{wrapfigure}\par
  \textbf{Onur Mutlu} is a Professor of Computer Science at ETH Zurich. He is
also a faculty member at Carnegie Mellon University, where he
previously held the Strecker Early Career Professorship.  His current
broader research interests are in computer architecture, systems,
hardware security, and bioinformatics. A variety of techniques he,
along with his group and collaborators, has invented over the years
have influenced industry and have been employed in commercial
microprocessors and memory/storage systems. He obtained his PhD and MS
in ECE from the University of Texas at Austin and BS degrees in
Computer Engineering and Psychology from the University of Michigan,
Ann Arbor. He started the Computer Architecture Group at Microsoft
Research (2006-2009), and held various product and research positions
at Intel Corporation, Advanced Micro Devices, VMware, and Google.  He
received the inaugural IEEE Computer Society Young Computer Architect
Award, the inaugural Intel Early Career Faculty Award, US National
Science Foundation CAREER Award, Carnegie Mellon University Ladd
Research Award, faculty partnership awards from various companies, and
a healthy number of best paper or "Top Pick" paper recognitions at
various computer systems, architecture, and hardware security
venues. He is an ACM Fellow "for contributions to computer
architecture research, especially in memory systems", IEEE Fellow for
"contributions to computer architecture research and practice", and an
elected member of the Academy of Europe (Academia Europaea). His
computer architecture and digital circuit design course lectures and
materials are freely available on YouTube, and his research group
makes a wide variety of software and hardware artifacts freely
available online. For more information, please see his webpage at
https://people.inf.ethz.ch/omutlu/.\par

 \begin{wrapfigure}{l}{25mm} 
    \includegraphics[width=1in,height=1.25in,clip,keepaspectratio]{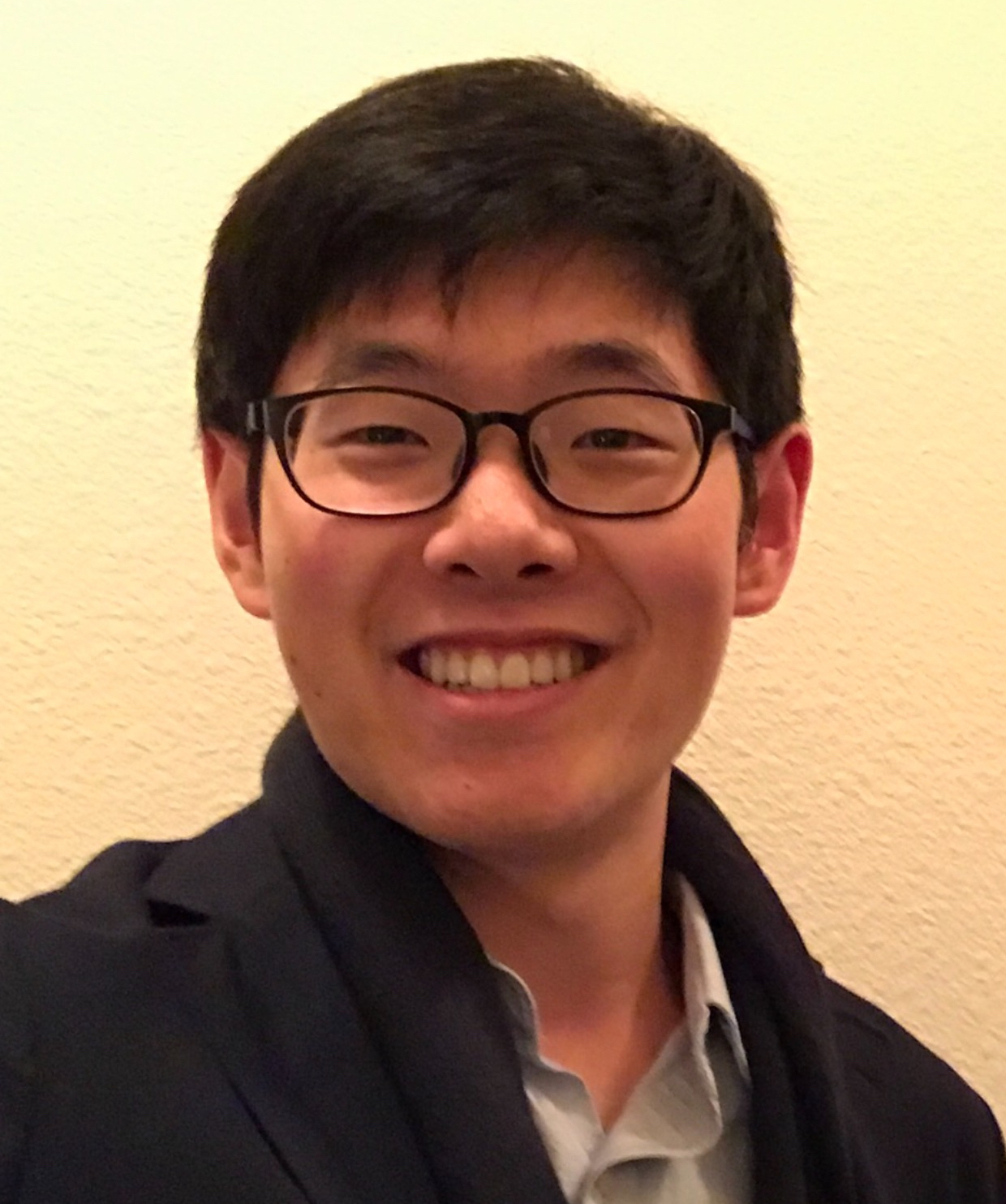}
  \end{wrapfigure}\par
  \textbf{Jeremie S. Kim} received the BS and MS degrees in Electrical and Computer Engineering from Carnegie Mellon University in Pittsburgh, Pennsylvania, in 2015. \jktwo{He is currently working on his PhD with Onur Mutlu at Carnegie Mellon University and ETH Zurich.} His current research interests are in computer architecture, memory latency/power/reliability, hardware security, and bioinformatics, and \jktwo{he has several publications on these topics}. \par

\end{document}

%% file: abstract.tex
\begin{abstract}

This retrospective paper describes the RowHammer problem in Dynamic
Random Access Memory (DRAM), which was initially introduced by Kim et
al. at the ISCA 2014 conference~\cite{rowhammer-isca2014}. RowHammer
is a prime (and perhaps the first) example of how a circuit-level
failure mechanism can cause a practical and widespread system
security vulnerability. It is the phenomenon that repeatedly accessing
a row in a modern DRAM chip causes bit flips in physically-adjacent
rows at consistently predictable bit locations. RowHammer is caused by
a hardware failure mechanism called {\em DRAM disturbance errors},
which is a manifestation of circuit-level cell-to-cell interference in
a scaled memory technology.

Researchers from Google Project Zero demonstrated in 2015 that this
hardware failure mechanism can be effectively exploited by user-level
programs to gain kernel privileges on real systems. Many other
follow-up works demonstrated other practical attacks exploiting
RowHammer. In this article, we comprehensively survey the scientific
literature on RowHammer-based attacks as well as mitigation techniques
to prevent RowHammer. We also discuss what other related
vulnerabilities may be lurking in DRAM and other types of memories,
e.g., NAND flash memory or Phase Change Memory, that can potentially
threaten the foundations of secure systems, as the memory technologies
scale to higher densities. We conclude by describing and advocating a
principled approach to memory reliability and security research that
can enable us to better anticipate and prevent such vulnerabilities.

\end{abstract}

%% file: introduction.tex
\section{Introduction and Outline}

Memory is a key component of all modern computing systems, often
determining the overall performance, energy efficiency, and
reliability characteristics of the entire system. The push for
increasing the density of modern memory technologies via technology
scaling, which has resulted in higher capacity (i.e., density) memory
and storage at lower cost, has enabled large leaps in the performance
of modern computers~\cite{mutlu-imw13}. This positive trend is clearly
visible in especially the dominant main memory and solid-state storage
technologies of today, i.e., DRAM~\cite{tldram, lee2017design,
  kevinchang-sigmetrics16, kim2018solar, chang_Understanding2017} and
NAND flash memory~\cite{cai-date12, cai-iccd13, cai2017error},
respectively.  Unfortunately, the same push has also greatly decreased
the reliability of modern memory technologies, due to the increasingly
smaller memory cell size and increasingly smaller amount of charge
that is maintainable in the cell, which makes the memory cell much
more vulnerable to various failure mechanisms and noise and
interference sources, both in DRAM~\cite{dram-isca2013,
  rowhammer-isca2014, samira-sigmetrics14, kang-memforum2014,
  avatar-dsn15, khan-dsn16, memcon-cal16, memcon-micro17, onur-date17,
  patel2017reach} and NAND flash nemory~\cite{cai2017error,
  cai-date12, cai-date13, cai-iccd13, cai-hpca15, cai-hpca17,
  cai-iccd12, cai-itj2013, cai-dsn15, cai-sigmetrics14, yixin-jsac16,
  onur-date17, luo2018improving, luo2018heatwatch, cai2017errors,
  warm-msst15}.

As memory scales down to smaller technology nodes, new failure
mechanisms emerge that threaten its correct operation. If such failure
mechanisms are not anticipated and corrected, they can not only
degrade system reliability and availability but also, perhaps even
more importantly, open up new security vulnerabilities: a malicious
attacker can exploit the exposed failure mechanism to take over the
entire system. As such, new failure mechanisms in memory can become
practical and significant threats to system security.

In this article, we provide a retrospective of one such example
failure mechanism in DRAM, which was initially introduced by Kim et
al. at the ISCA 2014 conference~\cite{rowhammer-isca2014}. We provide
a description of the RowHammer problem and its implications by
summarizing our ISCA 2014 paper~\cite{rowhammer-isca2014}, describe
solutions proposed by our original work~\cite{rowhammer-isca2014},
comprehensively examine the many works that build on our original work
in various ways, e.g., by developing new security attacks, proposing
solutions, and analyzing RowHammer. What comes next in this section
provides a roadmap of the entire article.

In our ISCA 2014 paper~\cite{rowhammer-isca2014}, we introduce the
RowHammer problem in DRAM, which is a prime (and perhaps the first)
example of how a circuit-level failure mechanism can cause a
practical and widespread system security vulnerability.  RowHammer, as
it is now popularly referred to, is the phenomenon that repeatedly
accessing a row in a modern DRAM chip causes bit flips in
physically-adjacent rows at consistently predictable bit locations. It
is caused by a hardware failure mechanism called DRAM disturbance
errors, which is a manifestation of circuit-level cell-to-cell
interference in a scaled memory technology. We describe the RowHammer
problem and its root causes in Section~\ref{sec:rowhammer-problem}.

Inspired by our ISCA 2014 paper's fundamental findings, researchers
from Google Project Zero demonstrated in 2015 that this hardware
failure mechanism can be effectively exploited by user-level programs
to gain kernel privileges on real
systems~\cite{google-project-zero,google-rh-blackhat}. Tens of other
works since then demonstrated other practical attacks exploiting
RowHammer, e.g., ~\cite{cloudflops, dedup-est-machina, anotherflip,
  qiao2016new, bhattacharya2016curious, jang2017sgx, aga2017good,
  pessl2016drama, rowhammer-js, flip-feng-shui, drammer,
  glitch-vu,fournaris2017exploiting, poddebniak2018attacking, nethammer, throwhammer,
  tatar2018defeating, carre2018openssl,
  barenghi2018software, zhang2018triggering,
  bhattacharya2018advanced, cojocar19exploiting}. These include remote takeover of a server vulnerable to
RowHammer, takeover of a victim virtual machine by another virtual
machine running on the same system, takeover of a mobile device by a
malicious user-level application that requires no permissions,
takeover of a mobile system quickly by triggering RowHammer using a
mobile GPU, and takeover of a remote system by triggering RowHammer on
it through the Remote Direct Memory Access (RDMA)
protocol~\cite{rdma-consortium}. We describe the works that build on RowHammer
to develop new security attacks in Section~\ref{sec:related-exploits}.

Our ISCA 2014 paper rigorously and experimentally analyzes the
RowHammer problem and examines seven different solutions, multiple of
which are already employed in practice to prevent the security
vulnerabilities (e.g., increasing the memory refresh rate).  We
propose a low-cost solution, Probabilistic Adjacent Row Activation,
which provides a strong and configurable reliability and security
guarantee; a solution whose variants are being adopted by DRAM
manufacturers and memory controller designers~\jk{\cite{x210-rh-ss}}. We describe this
solution and the six other solutions of our original paper in
Section~\ref{sec:rowhammer-solutions}. Many other works build on our
original paper to propose and evaluate other solutions to RowHammer,
and we discuss them comprehensively in
Section~\ref{sec:related-defenses}.

Our ISCA 2014 paper leads to a new mindset that has enabled a renewed
interest in hardware security research: general-purpose hardware is fallible, in a very widespread manner, and this causes real security problems.  We believe the RowHammer problem will
become worse over time since DRAM cells are getting closer to each
other with technology scaling. Other similar vulnerabilities may also
be lurking in DRAM and other types of memories, e.g., NAND flash
memory or Phase Change Memory, that can potentially threaten the
foundations of secure systems, as the memory technologies scale to
higher densities. Our ISCA 2014 paper advocates a principled
system-memory co-design approach to memory reliability and security
research that can enable us to better anticipate and prevent such
vulnerabilities. We describe promising ongoing and future research
directions related to RowHammer (Section~\ref{sec:future}), including
the examination of other potential vulnerabilities in memory (in
Section~\ref{sec:future-other-problems}) and the use of a principled
approach to make memory more reliable and more secure (in
Section~\ref{sec:future-prevention}).



%% file: rowhammer_summary.tex
\section{The RowHammer Problem: A Summary}
\label{sec:rowhammer-problem}

Memory isolation is a key property of a reliable and secure computing
system. An access to one memory address should not have unintended
side effects on data stored in other addresses. However, as process
technology scales down to smaller dimensions, memory chips become more
vulnerable to {\em disturbance}, a phenomenon in which different
memory cells interfere with each others' operation. We have shown, in
our ISCA 2014 paper~\cite{rowhammer-isca2014}, the existence of {\em
  disturbance errors} in commodity DRAM chips that are sold and used
in the field. Repeatedly reading from the same address in DRAM
could corrupt data in nearby addresses. Specifically, when a DRAM row
is opened (i.e., activated) and closed (i.e., precharged) repeatedly
(i.e., {\em hammered}), enough times within a DRAM refresh interval,
one or more bits in physically-adjacent DRAM rows can be flipped to
the wrong value. This DRAM failure mode is now popularly called {\em
  RowHammer}~\cite{rowhammer-arxiv16,rowhammer-wikipedia,rh-discuss,rh-twitter,rowhammer-thirdio,anvil,rowhammer-js,google-project-zero,google-rh-blackhat,dedup-est-machina,flip-feng-shui,drammer}. Using
an FPGA-based experimental DRAM testing infrastructure, which we
originally developed for testing retention time issues in
DRAM~\cite{dram-isca2013},\footnote{This infrastructure is currently
  released to the public, and is described in detail in our HPCA 2017
  paper~\cite{softmc}. The infrastructure has enabled many
  studies~\cite{aldram, dram-isca2013, rowhammer-isca2014, softmc,
    kevinchang-sigmetrics16, memcon-cal16, samira-sigmetrics14,
    avatar-dsn15, khan-dsn16, divadram-arxiv16, lee2017design,
    vampire2018, chang_Understanding2017} into the 
  failure and performance characteristics of modern DRAM, which were
  previously not well understood.} we tested 129 DRAM modules
manufactured by three major manufacturers (A, B, C) in seven recent
years (2008--2014) and found that 110 of them exhibited RowHammer
errors, the earliest of which dates back to 2010. This is illustrated
in Figure~\ref{fig:errors_vs_date}, which shows the error rates we
found in all 129 modules we tested where modules are categorized based
on manufacturing date.\footnote{Test details and experimental setup,
  along with a listing of all modules and their characteristics, are
  reported in our original RowHammer paper~\cite{rowhammer-isca2014}.}
In particular, {\em all} DRAM modules from 2012--2013 were vulnerable
to RowHammer, indicating that RowHammer is a recent phenomenon
affecting more advanced process technology generations (as also
demonstrated repeatedly by various works that come after our ISCA 2014
paper~\cite{rowhammer-thirdio, pessl2016drama, aga2017good,
  aichinger2015ddr, cojocar19exploiting, drammer}).


\input{errors_vs_date}

\subsection{RowHammer Mechanisms}

In general, disturbance errors occur whenever there is a strong enough
interaction between two circuit components (e.g., capacitors,
transistors, wires) that should be isolated from each other. Depending
on which component interacts with which other component and also how
they interact, many different modes of disturbance are possible. 

Among them, our ISCA 2014 paper identifies one particular disturbance
mode that affects commodity DRAM chips from all three major
manufacturers. {\em When a wordline's voltage is toggled repeatedly,
  some cells in nearby rows leak charge at a much faster rate than
  others.} Such {\em vulnerable cells}, if disturbed enough times,
cannot retain enough charge for even 64{\em ms}, the time interval at
which they are refreshed. Ultimately, this leads to the cells losing
data and experiencing disturbance errors.

Without analyzing existing DRAM chips at the device-level, which is an
option not available for us, we cannot make definitive claims about
how a wordline interacts with nearby cells to increase their
leakiness. Our ISCA 2014 paper hypothesizes, based on past studies and
findings, that there may be three ways of interaction. At least two
major DRAM manufacturers have confirmed all three of these hypotheses
as potential causes of disturbance errors. First, changing the voltage
of a wordline could inject noise into an adjacent wordline through
{\em electromagnetic coupling}~\cite{chao09, min90, redeker02}. This
partially enables the adjacent row of access-transistors for a short
amount of time and facilitates the leakage of charge from vulnerable
cells. Thus, if \jk{a row is hammered} enough times to disturb such
vulnerable cells before they get refreshed, charge in such cells get
drained to a point that the original cell values are not recoverable
any more. Second, {\em bridges} are a well-known class of DRAM faults
in which conductive channels are formed between unrelated wires and/or
capacitors~\cite{alars05, alars06}. One study on embedded DRAM (eDRAM)
found that toggling a wordline could accelerate the flow of charge
between two bridged cells~\cite{huang12}. Third, it has been reported
that toggling a wordline for hundreds of hours can permanently damage
it by {\em hot-carrier injection}~\cite{chia10}. If some of the
hot-carriers are injected into the neighboring rows, this could modify
the amount of charge in their cells or alter the \jk{characteristics} of
their access-transistors to increase their leakiness.

Several recent works have tried to examine and model RowHammer at the
circuit level; we survey these works in
Section~\ref{sec:related-circuit}.


\subsection{User-Level RowHammer}

Our ISCA 2014 paper also demonstrates that a very simple user-level
program~\cite{rowhammer-isca2014,safari-rowhammer} can reliably and
consistently induce RowHammer errors in three commodity AMD and Intel
systems using vulnerable DRAM modules. We released the source code of
this program~\cite{safari-rowhammer}, which Google Project Zero later
enhanced~\cite{google-rowhammer-test}. Using our user-level RowHammer
test program, we showed that RowHammer errors violate two invariants
that memory should provide: (i) a read access should not modify data
at any address and (ii) a write access should modify data only at the
address that it is supposed to write to. As long as a row is
repeatedly opened, both read and write accesses can induce RowHammer
errors, all of which occur in rows other than the one that is being
accessed. Since different DRAM rows are mapped (via mechanisms in the system software and the memory controller) to different software pages, our user-level program could
reliably corrupt specific bits in pages belonging to other
programs. As a result, RowHammer errors can be exploited by a
malicious program to breach memory protection and compromise the
system. In fact, we hypothesized, in our ISCA 2014 paper, that our
user-level program, with some engineering effort, could be developed
into a {\em disturbance attack} that injects errors into other
programs, crashes the system, or hijacks control of the system. We
left such research for the future since our primary objective in our
ISCA 2014 paper was to understand and prevent RowHammer
errors~\cite{rowhammer-isca2014}.

\subsection{Characteristics of RowHammer}

Our ISCA 2014 paper~\cite{rowhammer-isca2014} provides a detailed
experimental analysis of various characteristics of RowHammer,
including its prevalence across DRAM chips, access pattern dependence,
data pattern dependence, temperature dependence, address correlation
between victim and aggressor memory rows, number of bits in a victim
row that flip due to RowHammer in an adjacent row, number of rows that
get affected due to RowHammer in an adjacent row, relationship of
RowHammer-vulnerable cells with leaky cells that need higher refresh
rates, repeatability of RowHammer errors, the fact that a memory row
is vulnerable to RowHammer on both adjacent wordlines, and real system
demonstration of RowHammer. We omit these analyses here \jk{in this}
retrospective and focus on security vulnerabilities and prevention of
RowHammer. We refer the reader to~\cite{rowhammer-isca2014} for a
rigorous treatment of the characteristics of the RowHammer phenomenon.

One of the key takeaways from our characterization is that
RowHammer-induced errors are predictably repeatable. In other words,
if a cell's value gets corrupted via RowHammer, the same cell's value
is very likely to get corrupted again via RowHammer. This
repeatability enables the construction of repeatable security attacks
in a controlled manner, which we briefly \jk{discuss} next and cover in
detail in Section~\ref{sec:related-exploits}.

\subsection{RowHammer as a Security Threat}

RowHammer exposes a {\em security threat} since it leads to a breach
of memory isolation, where accesses to one row (e.g., a user-level
memory page) modifies the data stored in another memory row (e.g., a
privileged operating system page). As indicated above, malicious
software can be written to take advantage of these disturbance
errors. We call these {\em disturbance
  attacks}~\cite{rowhammer-isca2014}, or {\em RowHammer attacks}. Such
attacks can be used to corrupt system memory, crash a system, or take
over the entire system. Confirming the predictions of our ISCA
paper~\cite{rowhammer-isca2014}, researchers from Google Project Zero
developed a user-level attack that exploits RowHammer to gain kernel
privileges and thus take over an entire
system~\cite{google-project-zero, google-rh-blackhat}. More recently,
researchers showed that RowHammer can be exploited in various ways to
take over various classes of systems. As such, the RowHammer problem
has widespread and profound real implications on system security,
threatening the foundations of memory isolation on top of which modern
system security principles are built. We survey the \jk{works that
exploit} RowHammer to build many different security attacks in
Section~\ref{sec:related-exploits}.


\subsection{RowHammer Solutions}
\label{sec:rowhammer-solutions}

Our ISCA 2014 paper discusses and analyzes seven different
countermeasures to the RowHammer problem. Each solution makes a
different trade-off between feasibility, cost, performance, power, and
reliability. Among them, we believe our seventh and last solution,
called PARA, to be the most efficient \jk{with the lowest overhead.}

The first six solutions are: 1) \jk{manufacturing} better DRAM chips that are not
vulnerable, 2) using (strong) error correcting codes (ECC) to correct
RowHammer-induced errors, 3) increasing the refresh rate for all of
memory, \jktwo{4) statically remapping/retiring RowHammer-prone cells via a one-time post-manufacturing analysis, 5) dynamically remapping/retiring RowHammer-prone cells during system operation,} 6) accurately identifying hammered rows during
runtime and refreshing their neighbors.\footnote{Several early patent
  applications propose to maintain an array of counters (``detection
  logic'') in either the memory controller~\cite{bains14a, bains14b,
    greenfield14a} or in the DRAM chips themselves~\cite{bains14c,
    bains14d, greenfield14b}. If the counters are tagged with the
  addresses of only the most recently activated rows, \jk{the number of required counters} can
  be significantly reduced~\cite{greenfield14a}.} We will not go into
significant detail in this summary and retrospective, but none of
these first six solutions are very desirable as they come at
significant power, performance or cost overheads, as we describe in
our original work~\cite{rowhammer-isca2014}. We will revisit some of
these solutions in Section~\ref{sec:related-defenses} of this
\jk{article}, when we survey related work that builds on RowHammer.


Our ISCA 2014 paper's main proposal to prevent RowHammer is a
low-overhead mechanism called {\em PARA} ({\em probabilistic adjacent
  row activation}). The key idea of PARA is simple: every time a row
is opened and closed, one or more of its adjacent rows are also opened (i.e., refreshed) with some low probability $p$ (by the memory
controller or the DRAM chip). If one particular row happens to be
opened and closed repeatedly, then it is statistically certain that
the row's adjacent rows will eventually be opened as well, as we show
in our original work, assuming $p$ is chosen intelligently and
carefully. The main advantages of PARA are that 1) it is {\em
  stateless} in the sense that it does {\em not} require expensive
hardware data-structures to count the number of times that rows have
been opened or to store the addresses of the aggressor/victim rows, 2)
its performance and power consumption overheads are very low due to
the infrequent activation of {\em only adjacent rows} of a closed
row. Our ISCA 2014 paper provides a memory-controller-based implementation of PARA, evaluates
its reliability guarantee against adversarial access patterns, and
empirically examines its performance overhead. We show that by setting
the probability of refresh of adjacent rows $p$ to a reasonable yet
very low value (e.g., 0.001 or 0.005), PARA provides a strong
guarantee against RowHammer and leads to a very small performance
overhead of less than 0.75\%. More detailed discussion of the
implementation and evaluation of PARA can be found in our original
work. We will revisit PARA in Section~\ref{sec:related-defenses} of
this article.

%% file: errors_vs_date.tex
\begin{figure}[h]
\centering
\includegraphics{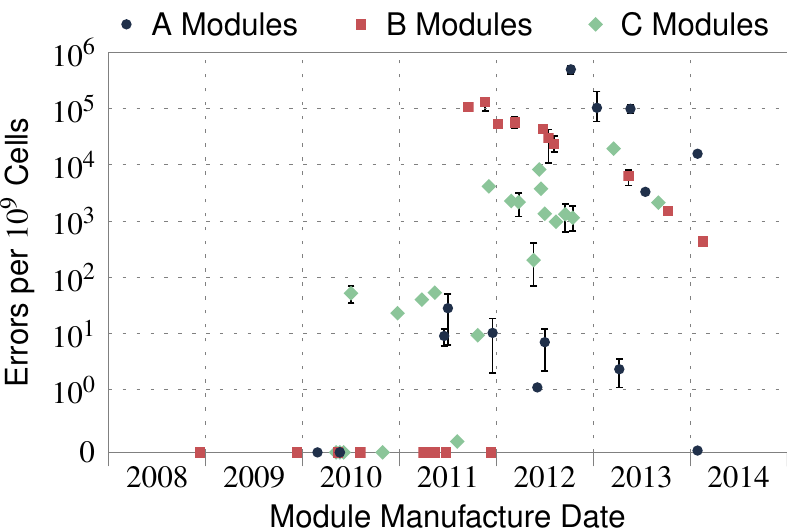}
\caption{RowHammer error rate vs.~manufacturing dates of 129 DRAM modules we tested (reproduced from~\cite{rowhammer-isca2014}).}
\label{fig:errors_vs_date}
\end{figure}

%% file: related.tex
\section{Survey of Works That Build on RowHammer}
\label{sec:related}

\ignore{
RowHammer exposes a {\em security threat} since it leads to a serious breach of
memory isolation: an access to one memory row (e.g., an OS page) predictably
modifies the data stored in another row (e.g., another OS page). Malicious
software, which we call {\em disturbance attacks}~\cite{rowhammer-isca2014}, or
{\em RowHammer attacks}, can be written to take advantage of these disturbance
errors to take over an entire system.
}

RowHammer has spurred \jk{a} significant amount of research since its
publication in 2014. In this section, we provide a categorical survey
of the array of works that build off of our original paper that
introduces the concept of RowHammer and {\em disturbance
  attacks}~\cite{rowhammer-isca2014}. We describe seven different
types of works: (1) security attacks that exploit the RowHammer
vulnerability, (2) defense and mitigation mechanisms against the
RowHammer phenomenon and the security attacks, (3) circuit-level
studies that aim to understand and model the RowHammer phenomenon, (4)
other works that exploit RowHammer for various purposes, (5) works
that build platforms to study RowHammer, (6) pop culture references to
RowHammer, and (7) works that show that \jk{the} RowHammer phenomenon continues
to exist in future generation DRAM chips younger than the ones we
examined in our original ISCA 2014 paper. While we describe the works,
we also point out the potential for future research in each topic area.

\subsection{Exploits using RowHammer} 
\label{sec:related-exploits}

Inspired by our ISCA 2014 paper's fundamental findings, researchers
from Google Project Zero demonstrated in 2015 that RowHammer can be
effectively exploited by user-level programs to gain kernel privileges
on real systems~\cite{google-project-zero,
  google-rh-blackhat}. Google Project Zero presented two exploits
using RowHammer. The first exploit runs as a Native Client (NaCl)
program and escalates privilege to escape from the x86-64 sandbox
environment. Since NaCl statically validates code before running it,
Google Project Zero simply shows that an attacker can modify safe
instructions to become unsafe. The second exploit, which is even more
powerful, runs as a normal x86-64 process on Linux and escalates
privilege to gain access to all of physical memory and thus take over
the entire system. The attacker hammers a page table entry (PTE) such
that it changes the PTE to point to a page table owned by the
attacking process. This gives the attacking process full read-write
access to its own page table and hence to all of physical memory,
which enables the attacking process to take over the entire system.

Tens of other works since then demonstrated other attacks exploiting
RowHammer \jk{and we explain several of them in \jktwo{some} detail here.  One involves} the
takeover of a victim virtual machine (VM) by
another attacker VM running on the same
system~\cite{flip-feng-shui}. In~\cite{flip-feng-shui}, the attacker
VM writes a memory page that it knows exists in the victim VM at a
RowHammer-vulnerable memory location. If memory deduplication merges
the victim VM's and attacker VM's duplicate pages to the attacker VM
page's location, the attacker \jk{can then induce RowHammer failures in the
deduplicated page's data, which is} shared by both the attacker and the victim. Since
RowHammer attacks modify memory without writes, the deduplication
engine does not detect the modification to memory, and the victim VM
continues to use the corrupted page. The authors show two attacks
using this method. The first attack compromises
OpenSSH~\cite{openssh} by modifying the public keys in a
victim VM such that the attacker can easily generate a private key
that matches the modified public key. It is easier to generate a
private key when a public key becomes easily factorable. The second
attack compromises the Linux package installation tool, {\em
  apt-get}~\cite{aptget} using two steps. First, the
attacker flips a bit in the apt-get domain name of the victim, such
that the victim's apt-get requests are redirected to a malicious
repository. Second, the attacker flips a bit in the page containing
the Ubuntu Archive Signing Keys, which are used to check the validity
of packages before installation. Thus, this work exploits the
RowHammer vulnerability to break both OpenSSH public key
authentication and install malicious software via widely-used
installation tools.

The Drammer work~\cite{drammer} demonstrates an attack that exploits
RowHammer on a mobile device using a malicious user-level application
that requires no permissions. This is the first demonstration of
RowHammer attacks on ARM-based systems. The work takes advantage of
the deterministic memory allocation patterns in the Android Linux
Operating System.  By exploiting these deterministic memory allocation
patterns, the authors present a methodology for forcing a victim
process to allocate its page table entry in a RowHammer-vulnerable
region of memory. To do this, the attacker process must essentially
allocate all possible memory regions for a page table allocation and
then release the page table allocation that contains the
RowHammer-vulnerable DRAM cells at bit offsets that enable
exploitation.  Because of the use of Buddy Allocation~\jk{\cite{knowlton1965fast}} (an allocation
scheme that forces allocations to the smallest available contiguous
region of memory) in Linux platforms, the attacker does not need to
allocate all of memory and risk crashing the system. The researchers
found 18 out of 27 phone models to be vulnerable to RowHammer and have
since released a \jktwo{mobile application} \jk{that tests memory for RowHammer-vulnerable cells and aggregates statistics on} how widespread the RowHammer
phenomenon is on mobile devices. This work shows that existing mobile
systems are widely vulnerable to RowHammer attacks.

\cite{rowhammer-js} demonstrates a remote takeover of a server vulnerable to
RowHammer via JavaScript code execution. Since JavaScript is present and
enabled by default in every modern browser, this work demonstrates the
proof-of-concept that the RowHammer attack can be launched by a website to
gain root privileges on a system that visits the website. 

Other works that have demonstrated attacks exploiting RowHammer
include takeover of a mobile system by triggering RowHammer using the
WebGL interface on a mobile GPU~\cite{glitch-vu, anotherflip},
takeover of a remote system by triggering RowHammer through the Remote
Direct Memory Access (RDMA) protocol~\cite{throwhammer, nethammer},
and various other attacks~\cite{cloudflops, dedup-est-machina,
  anotherflip, qiao2016new, bhattacharya2016curious, jang2017sgx,
  poddebniak2018attacking, aga2017good, tatar2018defeating,
  pessl2016drama, carre2018openssl, cojocar19exploiting,
  barenghi2018software, zhang2018triggering, bhattacharya2018advanced,
  fournaris2017exploiting}.  Thus, RowHammer has widespread and
profound real implications on system security, as it \jk{breaks} memory
isolation on top of which modern system security principles are built.

Our work has inspired many researchers to exploit RowHammer to devise
new attacks. As mentioned earlier, tens of papers were written in top
security venues that demonstrate various practical attacks exploiting
RowHammer (e.g.,~\cite{cloudflops, dedup-est-machina, anotherflip,
  qiao2016new, bhattacharya2016curious, jang2017sgx, aga2017good,
  pessl2016drama, rowhammer-js, flip-feng-shui, drammer,
  glitch-vu,fournaris2017exploiting, poddebniak2018attacking, nethammer, throwhammer,
  tatar2018defeating, carre2018openssl,
  barenghi2018software, zhang2018triggering,
  bhattacharya2018advanced, cojocar19exploiting, google-project-zero,
  google-rh-blackhat}).  These attacks started with
Google Project Zero's first work in 2015~\cite{google-project-zero,
  google-rh-blackhat} and they continue to this date, with the latest
ones that we know of being published in late
2018~\cite{poddebniak2018attacking, nethammer, throwhammer,
  tatar2018defeating, carre2018openssl,
  barenghi2018software, zhang2018triggering,
  bhattacharya2018advanced} \jk{and mid 2019~\cite{cojocar19exploiting}}. We believe there is a lot more to come
in this direction: as systems security researchers understand more
about RowHammer, and as the RowHammer phenomenon continues to
fundamentally affect memory chips due to technology scaling
problems~\cite{onur-date17}, researchers and practitioners will
develop different types of attacks to exploit RowHammer in various
contexts and in many more creative ways. Various recent reports
suggest that new-generation DDR4 DRAM and other DRAM chips are
vulnerable to RowHammer~\cite{rowhammer-thirdio, pessl2016drama,
  aga2017good, aichinger2015ddr, cojocar19exploiting}, as we examine
further in Section~\ref{sec:related-persistence}, so the fundamental
security research on RowHammer is likely to continue into the future.


\subsection{Defenses against RowHammer} 
\label{sec:related-defenses}


Our work also inspired many solution and mitigation techniques for
RowHammer from both researchers and industry practitioners. {\em
  Apple} publicly mentioned, in their critical security release for
RowHammer, that they increased the memory refresh rates due to the
``original research by Yoongu Kim et al. (2014)''~\cite{rh-apple}. The
industry-standard {\em Memtest86} program, which is used to test
deployed memory chips for errors, was updated, including a RowHammer
test, acknowledging our ISCA 2014 paper~\cite{rh-passmark}. Many
academic works developed solutions to RowHammer, working from our
original research (e.g.,~\cite{anvil, moin-rowhammer,
  seyedzadeh2017counter, brasser2016can, irazoqui2016mascat,
  son2017making, gomez2016dram, van2018guardion, lee2018twice,
  bu2018srasa, oh2018reliable}). Additionally, many patents for solutions to RowHammer
have been filed~\cite{bains2015row, bains14d, bains14c, bains14a,
  bains14b, greenfield14a}. We believe such solutions will continue to
be generated in both academia and industry, extending RowHammer's
impact into the very long term. We cover some of these solutions in
this section.

Given that RowHammer is such a critical vulnerability, it is important
to find both {\em immediate} and {\em long-term} solutions to the
RowHammer problem (as well as related problems that might cause
similar vulnerabilities). The goal of the immediate solutions is to
ensure that existing systems are patched such that the vulnerable DRAM
devices that are already in the field cannot be exploited. The goal of
the long-term solutions is to ensure that future DRAM devices do not
suffer from the RowHammer problem when they are released into the
field.

Given that immediate solutions require mechanisms that already exist
in systems operating in the field, they are fundamentally more
limited. A popular immediate solution, described and analyzed by our
ISCA 2014 paper~\cite{rowhammer-isca2014}, is to increase the refresh
rate of memory such that the probability of inducing a RowHammer error
before DRAM cells get refreshed is reduced. Several major system
manufacturers (including Apple, HP, Cisco, Lenovo, and IBM) have
adopted this solution and released security patches that increased
DRAM refresh rates (e.g.,~\cite{rh-apple,rh-hp,rh-lenovo,rh-cisco}) in
the memory controllers.  While this solution might be practical and
effective in reducing the vulnerability, it has the significant
drawbacks of increasing energy/power consumption, reducing system
performance, and degrading quality of service experienced by user
programs. Our paper shows that the refresh rate needs to be increased
by 7.8X its nominal value today, if we want to eliminate {\em all}
RowHammer-induced errors we saw in our tests of 129 DRAM modules!
Figure~\ref{fig:errors_vs_trefi} demonstrates this study: if we examine the most RowHammer-vulnerable module that we test from each manufacturer A, B, C, we find that completely eliminating the RowHammer-induced errors requires us to \jk{reduce} the refresh interval from the nominal 64ms to
8.2ms, leading to a 7.8X increase in the refresh rate. Since DRAM
refresh is already a significant
burden~\cite{raidr,darp-hpca2014,kang-memforum2014,samira-sigmetrics14,avatar-dsn15,patel2017reach}
on energy consumption, performance, and quality of service, increasing
it by any significant amount would only exacerbate the problem. Yet,
increased refresh rate is likely the most practical {\em immediate}
solution to RowHammer that does not require any significant change to
the system.

\input{errors_vs_trefi}

Other immediate solutions modify the software~\cite{konoth2018zebram,
  izzo2017reliably, brasser2016can, irazoqui2016mascat,
  google-project-zero, van2018guardion, wu2019protecting, oh2018reliable}. For example, ANVIL proposes
software-based detection of RowHammer attacks by monitoring via
hardware performance counters and selective explicit refreshing of
victim rows that are found to be under attack~\cite{anvil}.  One
short-term approach to mitigating RowHammer attacks is intelligently
allocating and physically isolating pages such that RowHammer cannot
affect important pages~\cite{brasser2016can, van2018guardion,
  konoth2018zebram}.  \cite{brasser2016can} extends the physical
memory allocator of the Operating System to allocate memory in such a
way that isolates memory pages of different system
entities. \cite{van2018guardion} prevents DMA-based attacks by
isolating DMA buffers with additional buffer rows (i.e., guard rows)
that do not store data. This ensures that any DMA-based attack can
only induce RowHammer bit flips in the guard rows without affecting
rows containing important data.  Another approach for mitigating
RowHammer statically analyzes code to identify code segments that are
probably RowHammer attacks and prevents them prior to
execution~\cite{irazoqui2016mascat}.  ZebRAM~\cite{konoth2018zebram}
reserves odd rows as ``safe'' rows and even rows as ``unsafe'' rows
such that hammering a safe row should never result in a RowHammer
failure in another safe row. The unsafe rows are used as swap space
and a portion of safe rows are used as a cache for data in the
unsafe-row swap space.  Whenever data in unsafe rows is migrated to
safe rows, ZebRAM performs software integrity checks and error
correction.  Unfortunately, such software-based solutions usually 1)
require modifications to system software, 2) might be intrusive to
system operation, and/or 3) might cause significant performance or
memory space overheads (yet they are still promising to research).

As briefly discussed in Section~\ref{sec:rowhammer-solutions}, our
ISCA 2014 paper~\cite{rowhammer-isca2014} discusses and analyzes seven
short-term and long-term countermeasures to the RowHammer problem. The
first six solutions are: 1) making better DRAM chips that are not
vulnerable, 2) using (strong) error correcting codes (ECC) to correct
RowHammer-induced errors, 3) increasing the refresh rate for all of
memory, \jktwo{4) statically remapping/retiring RowHammer-prone cells via a one-time post-manufacturing analysis, 5) dynamically remapping/retiring RowHammer-prone cells during system operation,} 6) accurately identifying hammered rows during
runtime and refreshing their neighbors. Our work shows that the first
six solutions are not very desirable as they come at significant
power, performance or cost overheads. We already discussed the
overheads of increasing the refresh rates across the board. Similarly,
the use of simple SECDED (single-error correcting double-error
detecting) error correcting codes (ECC), as employed in many server
and datacenter systems, is {\em not} enough to prevent all RowHammer
errors, as some cache blocks experience two or more bit flips,
which are not correctable by SECDED ECC, as we have shown in our ISCA
2014 paper~\cite{rowhammer-isca2014}. Table~\ref{table:secded}
demonstrates this by showing how many 64-bit words in the full
address-space (0--2GB) of the most RowHammer-vulnerable DRAM modules
of the three major DRAM manufacturers contain 1, 2, 3, or 4 victim
cells. While most words have just a single victim, there are also some
words with multiple victims. Thus, stronger ECC is very likely
required to correct RowHammer errors, which comes at the cost of
additional energy, performance, cost, and DRAM capacity
overheads.\footnote{Note that protecting {\em all} memory rows with strong ECC is likely a wasteful solution for RowHammer because RowHammer-induced bit-flips are access-pattern dependent; they are not randomly-occurring bit-flips. Since only a small number of rows can be hammered at a given time, paying the capacity, cost, and energy overheads of extra redundancy required for strong ECC for {\em all} memory rows, solely to protect them against RowHammer, is likely not an efficient solution to the RowHammer problem.}  Alternatively, the sixth solution described above, i.e.,
accurately identifying a row as a hammered row requires keeping track
of access counters for a large number of rows in the memory
controller~\cite{moin-rowhammer}, potentially leading to very large
hardware area and power consumption, and  performance,
overheads.

\input{secded.tex}

There are many other works that propose long-term
solutions~\cite{gomez2016dram, bu2018srasa, gong2018memory,
  jones2017holistic, kline2017sustainable, schilling2018pointing,
  seyedzadeh2017counter, lee2018twice, son2017making, danger2018ccfi,
  bains2015row, bains14d, greenfield14a, wang2019detect, vig2018rapid, kim2019effective}, building on our original
work.  \cite{son2017making} uses a probabilistic mechanism similar to
PARA in the original RowHammer paper in addition to a small stack for
maintaining access history information to determine whether adjacent
rows need to be refreshed to avoid bit flips.
\cite{seyedzadeh2017counter, lee2018twice, vig2018rapid, wang2019detect} are
counter-based defenses that rely on maintaining access counts to DRAM
rows and refreshing adjacent rows when the access count of a row exceeds a pre-determined threshold. 
These works focus on reducing the overhead of counting accesses to
DRAM addresses to enable a viable implementation of the sixth soution
we described above.

\ignore{
\cite{van2018guardion} prevents DMA-based attacks by
isolating DMA buffers with additional buffer rows (i.e., guard rows)
that do not store data. This ensures that any DMA-based attack can
only induce RowHammer bit flips in the guard rows without affecting
rows containing important data.
}

We believe the long-term solution to RowHammer can actually be very
simple and low cost: when the memory controller closes a row (after it
was activated), it, with a very low probability, refreshes the
adjacent rows. The probability value is a parameter determined by the
system designer or provided programmatically, if needed, to trade off
between performance overhead and vulnerability protection
guarantees. We show that this probabilistic solution, called {\em PARA
  (Probabilistic Adjacent Row Activation)}, is extremely effective: it
eliminates the RowHammer vulnerability, providing much higher
reliability guarantees than modern hard disks today, while requiring
no storage cost and having negligible performance and energy
overheads~\cite{rowhammer-isca2014}.

PARA is {\em not} immediately implementable in an existing system
because it requires changes to either the memory controllers or the
DRAM chips, depending on where it is implemented. If PARA is
implemented in the memory controller, the memory controller needs to
obtain information on which rows are adjacent to each other in a DRAM
bank.  This information is currently {\em unknown} to the memory
controller as DRAM manufacturers can internally remap rows to other
locations~\cite{dram-isca2013,rowhammer-isca2014,khan-dsn16,memcon-cal16,lee2017design}
for various reasons, including for tolerating various types of
faults. However, this information can be simply provided by the DRAM
chip to the memory controller using the serial presence detect (SPD)
read-only memory present in modern DRAM modules, as described in our
ISCA 2014 paper~\cite{rowhammer-isca2014}. It appears that some very
recent Intel memory controllers implement a limited variant of PARA,
whose adjacent-row activation probability can be chosen by the user
via modifications in the BIOS~\cite{x210-rh-ss}. If PARA is
implemented in the DRAM chip, then the hardware interface to the DRAM
chip should be such that it allows DRAM-internal refresh operations
that are not initiated by an external memory controller. This could be
achieved with the addition of a new DRAM command, like the {\em
  targeted refresh} command proposed in a patent by
Intel~\cite{intel-rh-refresh}. In 3D-stacked memory
technologies~\cite{ramulator,smla-taco16}, e.g., HBM (High Bandwidth
Memory)~\cite{hbm,smla-taco16} or HMC (Hybrid Memory Cube)~\cite{hmc},
which combine logic and memory in a tightly integrated fashion, the
logic layer can be easily modified to implement
PARA.\footnote{Alternatively, for a solution like PARA to be
  implemented in the DRAM chip, without modifying the hardware
  interface to the DRAM chip, one can exploit the timing slack in the
  DRAM timing parameters that already exist under various
  conditions. For example, the timing slack in the specified precharge
  timing parameter or the refresh latency parameter can be exploited
  by the DRAM chip itself to internally issue refresh operations to
  targeted rows with some probability. Even though such timing slack
  exists in DRAM chips, as shown by many recent experimental
  studies~\cite{aldram,kevinchang-sigmetrics16,lee2017design,kim2018solar,vrl-dram},
  we do not believe this is a robust solution since 1) the timing
  slack may not exist under all operating conditions or for all chips,
  2) many studies would like to reduce the timing slack as much as
  possible to improve DRAM performance and
  energy~\cite{aldram,kevinchang-sigmetrics16,lee2017design,kim2018solar,vrl-dram}.} Alternatively, if the memory interface is asynchronous with the processor, a simple controller that is tightly coupled with the memory chip can freely and easily implement PARA internally to the memory chip.

All these implementations of the promising PARA solution are examples
of much better cooperation between memory controller and the DRAM
chips.  Regardless of the exact implementation, we believe RowHammer,
and other upcoming reliability vulnerabilities like RowHammer, can be
much more easily found, mitigated, and prevented with better
cooperation between and co-design of system and memory, i.e., {\em
  system-memory co-design}~\cite{mutlu-imw13}. System-memory co-design
is explored by recent works for mitigating various DRAM-based security
and DRAM scaling issues, including retention failures and performance
problems (e.g.,~\cite{raidr, salp, mutlu-imw13, kang-memforum2014,
  superfri14, hrm-dsn2014, samira-sigmetrics14, avatar-dsn15,
  khan-dsn16, memcon-cal16, aldram, tldram, rowclone, lisa,
  kevinchang-sigmetrics16, kim2018solar, dram-isca2013, darp-hpca2014,
  rowhammer-isca2014, lee2017design, chargecache-hpca16, gs-dram,
  ddma-pact15, vivek-and-or, kim2018dram, kim2019drange,
  keller2014dynamic, tehranipoor2015dram, sutar2018d, xiong2016run,
  sutar2016d, rahmati2015probable, tang2017dram,
  tehranipoor2017investigation}).  Taking the system-memory co-design
approach further, providing more intelligence and
configurability/programmability/patch-ability in the memory controller can greatly
ease the tolerance to errors like RowHammer: when a new failure
mechanism in memory is discovered, the memory controller can be
configured/programmed/patched to execute specialized functions to
profile and correct for such mechanisms. We believe this direction is
very promising, and several works have explored {\em online profiling}
mechanisms for fixing retention
errors~\cite{samira-sigmetrics14,avatar-dsn15,khan-dsn16,
  memcon-cal16, memcon-micro17, patel2017reach}, reducing
latency~\cite{lee2017design}, and reducing energy
consumption~\cite{chang_Understanding2017}.  These works provide
examples of how an intelligent memory controller can alleviate the
retention failures, and thus the DRAM refresh problem~\cite{raidr,
  dram-isca2013}, as well as the DRAM latency problem~\cite{tldram,
  aldram}.

\subsection{Circuit-level Studies of RowHammer} 
\label{sec:related-circuit}

A very recent work~\cite{yang2019trap} presents evidence via 3D CAD
simulations with single charge traps, that the RowHammer effect is
governed by the charge pumping process. The RowHammer effect is
exacerbated when charge is captured around an aggressor wordline and
carriers migrate to victim wordlines. The authors also find that
feature size scaling aggravates the RowHammer effect, which could make
it more difficult to mitigate in future DRAM generations.

\cite{yun2018study} provides a study of the effects of irradiating DRAM on the RowHammer phenomenon, with two major  findings. First, the study finds that irradiating DRAM with gamma rays
increases the number of DRAM rows that are vulnerable to
RowHammer. Second, the authors correlate the cells that are
vulnerable to RowHammer with those that have low data retention times
and they find almost no correlation, corroborating the results of our
ISCA 2014 paper.  

\cite{lim2017active} also irradiates DRAM with gamma rays, which results in cells with lower data retention times and cells with a higher susceptibility to RowHammer failures. The authors then perform \emph{temperature annealing} (i.e., a method for baking DRAM at a high temperature to ``repair'' retention-weak cells) on the DRAM devices and find that cells that experience a higher susceptibility to RowHammer after irradiation maintain the higher
susceptibility to RowHammer even after temperature annealing.  

\cite{ryu2017overcoming} presents evidence that hydrogen (H2)
annealing of cell-transistors during the dry etch process shows a
reduction in interface trap density. Since the RowHammer failure is
mainly caused by the traps in the interface (according to the authors'
hypotheses), the authors show that this technique can help to improve
DRAM reliability against crosstalk and thus alleviate RowHammer attacks.

\jktwo{\cite{park2016experiments} and \cite{park2016statistical} experimentally
test DDR3 devices for RowHammer susceptibility, show statistical distributions of RowHammer failures across many devices, and present evidence that \jkthree{the root cause of the RowHammer phenomenon is charge recombination of the victim cell \jkfour{with electrons from the current channels between neighboring cells and their corresponding bitlines}}.}


\subsection{Other Works Exploiting RowHammer}
\label{sec:related-other}

There are other papers that build upon RowHammer but do not
necessarily show a RowHammer attack or defense.  One work shows that
the RowHammer phenomenon can be used as a security
primitive. \cite{schaller2017intrinsic} shows that RowHammer can be
used as an effective Physical Unclonable Function (PUF), a function
that generates unique identifiers (i.e., fingerprints) of a device
based on the unique properties of the device due to manufacturing
variation. The authors experimentally show that by reserving a region
of memory and inducing RowHammer failures in each of the rows of the
region, they can generate bit flips in the region whose locations are
unique to the device and can be used to identify the device. A more
recent work~\cite{zeitouni2018s} presents an attack on the
RowHammer-based PUF~\cite{schaller2017intrinsic} by effectively
showing that hammering on rows surrounding the region reserved by the
RowHammer-based PUF causes the rows at the edges of the reserved DRAM
region to have an increased number of bit flips. This results in a
modification of the resulting fingerprint, which then results in an
unidentifiable device.

\subsection{Platforms for Studying RowHammer} 
\label{sec:related-platforms}

Many prior works present ways to make studying RowHammer easier.
\cite{francis2018raspberry} describes their Raspberry Pi Operating System for
exploring memory concepts simply due to a direct linear mapping between virtual
addressses to physical addresses. This mitigates the difficulty of determining
which DRAM rows are physically adjacent. SoftMC~\cite{softmc,
softmc-safarigithub} is an FPGA-based memory controller implementation that
enables testing custom DRAM timing parameter values with direct access to DRAM
physical addresses.  Drammer~\cite{drammer-github, drammer-app-github} is an
open-source Android application that tests mobile devices for vulnerability to
the RowHammer exploit and gathers data from users to determine how widespread
the RowHammer vulnerability is across many generations of mobile devices.
MemTest86~\cite{rh-passmark} is software that tests DRAM for many types of
reliability issues. As described above, after our original ISCA 2014 paper,
MemTest86 developers added RowHammer testing to their suite, which enables
users to test their system for the RowHammer vulnerability.
\cite{barenghi2018software, izzo2017reliably, oh2018reliable} provide methods for reverse
engineering DRAM address mapping such that attackers can determine the two rows
that surround a victim row and \jk{hammer the victim row more effectively for causing RowHammer failures}.
\cite{vila2019theory} provides an algorithm for determining the eviction set of
cache lines in linear time such that an attacker can maximize accesses to DRAM
even when caching is unavoidable.  \cite{aichinger2015ddr} repurposes a DDR
protocol analyzer with a DIMM interposer to count the activations to each row
within a 64 ms interval to detect whether RowHammer occurs in any application.

\subsection{Media References to RowHammer} 
\label{sec:related-popculture}

Our ISCA 2014 work also turned RowHammer into a popular phenomenon
(e.g.,~\cite{rowhammer-wikipedia, rh-discuss, rh-twitter, rh-zdnet1,
  rowhammer-thirdio, google-rh-blackhat, rh-passmark, rh-futureplus,
  arstechnica_ddr4-rh, arstechnica_rh, wired-rh2, rh-techrepublic1,
  rh-zdnet4, rh-zdnet3, rh-zdnet2, arstechnica_rh3, rh-hackernews1,
  rh-hackernews2, rh-register1, bleepingcomputer-rh, cyware-rh1,
  threatpost-rh1}), which, in turn, \jk{has helped make hardware security \jktwo{even more} 
"mainstream" in popular media and the broader} security community.  It \jk{showed that hardware reliability problems can be very serious security threats that
have to be defended against.}
A well-read article
from the Wired magazine, all about RowHammer, is entitled ``Forget
Software -- Now Hackers are Exploiting Physics!''~\cite{wired-rh},
indicating the shift of mindset towards \jk{very low-level} hardware security
vulnerabilities in the popular mainstream security community.  Many
other popular articles in press have been written about RowHammer,
many of which \jk{pointing to our ISCA 2014 work~\jktwo{\cite{rowhammer-isca2014}} as the first demonstration 
and scientific analysis of the RowHammer problem. Showing that hardware
reliability problems can be serious security threats and pulling them to the
popular discussion space, and thus influencing the mainstream discourse},
creates a very long term impact for the RowHammer problem and thus our original ISCA 2014 paper.

\subsection{Persistence of RowHammer Failures in Modern DRAM} 
\label{sec:related-persistence}

Unfortunately, despite the many proposals in industry and academia to fix the
RowHammer issue, \jk{RowHammer failures} still seem to be observable in state-of-the-art DRAM devices
in a variety of generations and standards (e.g.,
DDR4~\cite{rowhammer-thirdio,pessl2016drama, aga2017good, aichinger2015ddr},
ECC DRAM~\cite{cojocar19exploiting}, LPDDR3 and LPDDR2 DRAM~\cite{drammer}). This
persisting phenomenon suggests that the security vulnerabilities might continue
in the current generation of DRAM chips as well.  As such, it is critical
to continue to investigate solutions to the RowHammer vulnerability.

\subsection{RowHammer in a Broader Context}

Springing off from the stir created by RowHammer, we take a step back
and argue that there is little that is surprising about the fact that
we are seeing disturbance errors in the heavily-scaled DRAM chips of
today. Disturbance errors are a general class of reliability problems
that is present in not only DRAM, but also other memory and storage
technologies. All scaled memory technologies, including
SRAM~\cite{chen05, guo09, kim11}, flash~\cite{cai-date12, cai-date13,
  cai-iccd13, cai-dsn15, cooke07, cai-hpca15, cai2017errors,
  cai2017error, luo2018improving, luo2018heatwatch,
  justin-flash-sigmetrics15, flash-field-analysis2}, and hard disk
drives~\cite{jiang03, tang08, wood09}, exhibit such disturbance
problems.  In fact, two of our works experimentally examine read disturb errors in flash memory: 1) our original work in DATE 2012~\cite{cai-date12} that provides a rigorous experimental study of error patterns in modern MLC NAND flash memory chips demonstrates the importance of read disturb error patterns, 2) our recent work at DSN 2015~\cite{cai-dsn15}
experimentally characterizes the read disturb errors in flash memory,
shows that the problem is widespread in recent flash memory chips, and
develops mechanisms to correct such errors in the flash memory
controller. Even though the mechanisms that cause the bit flips are
different in different technologies, the high-level root cause of the
problem, {\em cell-to-cell interference}, due to the fact that the
memory cells are too close to each other, is a fundamental issue that
appears and will likely continue to appear in any technology that scales down to small
enough technology nodes~\cite{cai2017errors,yang2019trap}. Thus, we
should expect such problems to continue as we scale any memory
technology, including emerging ones, to higher densities.

What sets DRAM disturbance errors apart from other technologies'
disturbance errors is that 1) DRAM is exposed to the user-level
programs and manipulated directly by a program's load and store
instructions (which we do not anticipate to change any time soon, since direct data
manipulation in main memory is a fundamental component of programming languages and systems), and 2)
in modern DRAM, as opposed to other technologies, strong error
correction mechanisms are {\em not} commonly employed (either in the
memory controller or the memory chip). The success of DRAM scaling
until recently has {\em not} relied on a memory controller that
corrects errors (other than performing periodic refresh and more
recently employing very simple single-error correcting
codes~\cite{kang-memforum2014, nair2013case, nair2016xed,
  micron2017whitepaper, oh20153, patel2019understanding}). Instead, DRAM chips were implicitly
assumed to be error-free and did {\em not} require the help of the
controller to operate correctly. Thus, such errors were perhaps not as
easily anticipated and corrected within the context of DRAM. In
contrast, the success of other technologies, e.g., flash memory and
hard disks, has heavily relied on the existence of an intelligent
controller that plays a key role in correcting errors and making up
for reliability problems of the memory chips themselves~\cite{cai2017error}.  This has not
only enabled the correct operation of assumed-faulty memory chips but
also enabled a mindset where the controllers are co-designed with the
chips themselves, covering up the memory technology's deficiencies and
hence perhaps enabling better anticipation of errors with technology
scaling. This approach is very prominent in modern SSDs (solid state
drives), for example, where the flash memory controller employs a wide
variety of error mitigation and correction
mechanisms~\cite{cai-iccd12, cai-date12, cai-date13, cai-iccd13,
  cai-sigmetrics14, cai-dsn15, cai-hpca17, cai-hpca15, cai-itj2013,
  yixin-jsac16, cai2017errors, cai2017error}, including not only
sophisticated strong ECC mechanisms but also targeted voltage
optimization, retention mitigation and disturbance mitigation
techniques. We believe changing the mindset in modern DRAM to a
similar mindset of {\em assumed-faulty memory chip and an intelligent
  memory controller that makes it operate correctly} can not only
enable better anticipation and correction of future issues like
RowHammer but also better scaling of DRAM into future technology
nodes~\cite{mutlu-imw13}.

%% file: errors_vs_trefi.tex
\begin{figure}[h]
\centering
\includegraphics{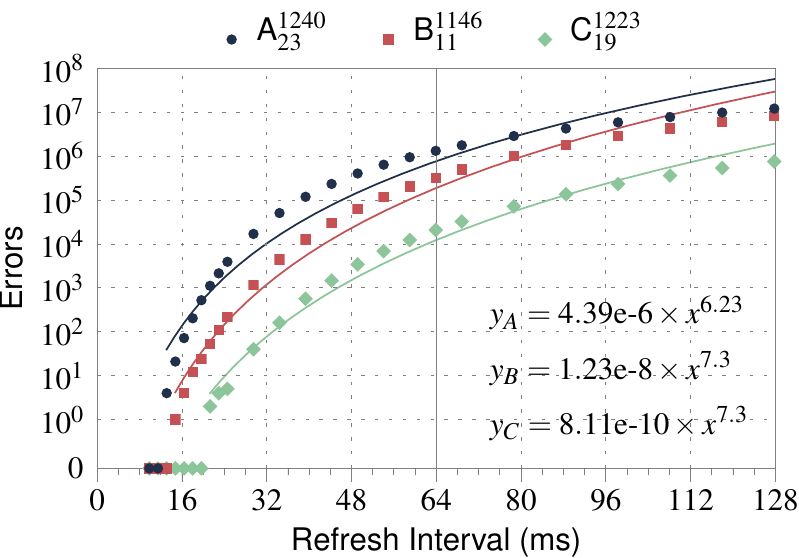}
\caption{Number of RowHammer-induced errors observed on the most RowHammer-vulnerable module of each DRAM manufacturer A, B, C, as the refresh interval is varied from 8ms to 128ms (reproduced from~\cite{rowhammer-isca2014}).}
\label{fig:errors_vs_trefi}
\end{figure}

%% file: secded.tex
\begin{table}[h]
\centering
\footnotesize
\sisetup{group-separator={,}}
\sisetup{group-minimum-digits=4}
\sisetup{detect-weight=true, detect-inline-weight=math}
\begin{tabular}{c*{4}{S[table-format=7]}}

\toprule

\multirow{2}{*}[-2.5pt]{\em \centering Module}    &
\multicolumn{4}{c}{\multirow{1}{*}{\em \centering Number of 64-bit words with X errors}}  \\

\cmidrule(lr){2-5}

   &   

\multicolumn{1}{r}{\textit{X} $=$ 1}   &   
\multicolumn{1}{r}{\textit{X} $=$ 2}   &   
\multicolumn{1}{r}{\textit{X} $=$ 3}   &   
\multicolumn{1}{r}{\textit{X} $=$ 4}   \\


\midrule

\module{A}{23}{}  &  9709721   &   \bfseries 181856  &  \bfseries 2248 & \bfseries 18 \\


\module{B}{11}{}  &  2632280   &  \bfseries 13638   & \textbf{47}      & {\centering 0} \\


\module{C}{19}{}  &  141821   &   \bfseries 42      &  {\centering 0}  & {\centering 0} \\

\bottomrule

\end{tabular}
\caption{Uncorrectable multi-bit RowHammer errors (in bold) observed on the most RowHammer-vulnerable module of each DRAM manufacturer A, B, C (reproduced from~\cite{rowhammer-isca2014})} 
\label{table:secded}
\end{table}

%% file: future.tex
\section{Ongoing and Future Work} 
\label{sec:future}

We believe there is a lot more research to come that will build on
RowHammer, from at least three perspectives: 1) the security attack
perspective, 2) the defense/mitigation perspective, 3) a broader
understanding, modeling, and prevention \jk{perspective}.

As systems security researchers understand more about RowHammer, and
as the RowHammer phenomenon continues to fundamentally affect memory
chips due to technology scaling problems~\cite{onur-date17},
researchers and practitioners will develop different types of attacks
to exploit RowHammer in various contexts and in many more creative
ways. RowHammer is a critical problem that manifests in the
difficulties in DRAM scaling and is expected to only become worse in
the future~\cite{mutlu13, mutlu2015main, superfri14}. As we discussed,
some recent reports suggest that new-generation DRAM chips are
vulnerable to RowHammer (e.g.,
DDR4~\cite{rowhammer-thirdio,pessl2016drama, aga2017good,
  aichinger2015ddr},
ECC~\cite{rowhammer-isca2014,cojocar19exploiting},
LPDDR3 and LPDDR2~\cite{drammer}). This indicates that effectively mitigating the
RowHammer problem with low overhead is difficult and becomes more
difficult as process technology scales further. Even with the wide
array of works that build on top of RowHammer, we believe that these
papers have yet to scratch the surface of this field of reliability
and security, especially as manufacturing technology scaling continues
in all technologies. It is critical to deeply understand the
underlying factors of the RowHammer problem (and more generally the
crosstalk problem) such that we can effectively prevent these issues
across all technologies with minimal overhead.  As DRAM cells become
even smaller and less reliable, it is likely for them to become even
more vulnerable to complicated and different modes of failure that
are sensitized only under specific access-patterns and/or
data-patterns. As a scalable solution for the future, our ISCA 2014
paper argues for adopting a system-level approach~\cite{mutlu-imw13}
to DRAM reliability and security, in which the DRAM chips, the memory
controller, and perhaps the operating system collaborate together to
diagnose/treat emerging DRAM failure modes.

We believe that more and more researchers will focus on providing
security in all aspects of computing so that such hardware faults that are exposed to the software (and thus the public) are minimized. RowHammer enabled a shift of mindset among mainstream security researchers: general-purpose
hardware is fallible (in a very widespread manner) and its problems
are actually exploitable. This shift of mindset enabled many systems
security researchers to examine hardware in more depth and understand
its inner workings and vulnerabilities better. We believe it is no
coincidence that two of the groups that concurrently discovered the
heavily-publicized Meltdown~\cite{lipp2018meltdown} and
Spectre~\cite{kocher2018spectre} vulnerabilities (Google Project Zero
and TU Graz InfoSec) have heavily worked on RowHammer attacks
before. We believe this shift in mindset, enabled in good part by the
existence and prevalence of RowHammer, will continue to be very be
important for discovering and solving other potential vulnerabilities
that may rise as a result of both technology scaling and hardware
design.

\subsection{Other Potential Vulnerabilities}
\label{sec:other-problems}
\label{sec:future-other-problems}

We believe that, as memory technologies scale to higher densities,
other problems may start appearing (or may already be going unnoticed)
that can potentially threaten the foundations of secure systems. There
have been recent large-scale field studies of memory errors showing
that both DRAM and NAND flash memory technologies are becoming less
reliable~\cite{superfri14,
  justin-memerrors-dsn15,dram-field-analysis2, dram-field-analysis3,
  dram-field-analysis4, justin-flash-sigmetrics15,
  flash-field-analysis2, cai2017errors, cai2017error,
  luo2018improving, luo2018heatwatch, cai-date12, cai-hpca15,
  mutlu-imw13, patel2017reach, onur-date17}. As detailed experimental
analyses of real DRAM and NAND flash chips show, both technologies are
becoming much more vulnerable to cell-to-cell interference
effects~\cite{superfri14, rowhammer-isca2014, cai-dsn15,
  cai-sigmetrics14, cai-iccd13, cai-date12,cai-date13, flash-fms-talk,
  yixin-jsac16, cai-hpca17, cai2017errors, cai2017error, mutlu-imw13,
  onur-date17}, data retention is becoming significantly more
difficult in both technologies~\cite{raidr,samira-sigmetrics14,
  dram-isca2013, khan-dsn16, avatar-dsn15, darp-hpca2014,
  kang-memforum2014, mandelman-jrd02, cai-hpca15, cai-iccd12,
  warm-msst15, cai-date12,cai-date13,cai-itj2013, flash-fms-talk,
  memcon-cal16, cai2017errors, cai2017error, luo2018improving,
  luo2018heatwatch, superfri14, mutlu-imw13}, and error variation
within and across chip, and across operating conditions, is increasingly prominent~\cite{dram-isca2013,
  aldram, kevinchang-sigmetrics16, dram-process-variation-3,
  cai-date12, cai-date13, lee2017design, kim2018solar, kim2018dram,
  kim2019drange}.  Emerging memory technologies~\cite{mutlu-imw13,
  meza-weed13}, such as Phase-Change Memory~\cite{pcm-isca09,
  zhou-isca09, moin-isca09, moin-micro09, wong-pcm, raoux-pcm,
  pcm-ieeemicro10, pcm-cacm10, justin-taco14, rbla},
STT-MRAM~\cite{chen-ieeetmag10,kultursay-ispass13}, and
RRAM/ReRAM/memristors~\cite{wong-rram} are likely to exhibit similar
and perhaps even more exacerbated reliability issues. We believe, if
not carefully accounted for and corrected, these reliability problems
may surface as security problems as well, as in the case of RowHammer,
especially if the technology is employed as part of the main memory
system that is directly exposed to user-level programs.

We briefly examine two example potential vulnerabilities. We believe
future work examining these vulnerabilities, among others, are
promising for both fixing the vulnerabilities and enabling the
effective scaling of memory technology.

\subsubsection{Data Retention Failures}

Data retention is a fundamental reliability problem, and hence a
potential vulnerability, in especially charge-based memories like DRAM and flash memory. This is because charge leaks out of the charge storage unit
(e.g., the DRAM capacitor or the NAND flash floating gate) over
time. As such memories become denser, three major trends make data
retention more difficult~\cite{raidr,dram-isca2013,kang-memforum2014,
cai-hpca15}. First, the number of memory cells increases, leading to
the need for more refresh operations to maintain data
correctly. Second, the charge storage unit (e.g., the DRAM capacitor)
becomes smaller and/or morphs in structure, leading to potentially
lower retention times. Third, the voltage margins that separate one
data value from another become smaller (e.g., the same voltage window
gets divided into more ``states'' in NAND flash memory, to store more
bits per cell), and, as a result, the same amount of charge loss is more
likely to cause a bit error in a smaller technology node than in a larger
one.

\textbf{DRAM Data Retention Issues}

Data retention issues in DRAM are a fundamental scaling limiter of the
DRAM technology~\cite{mutlu-imw13, dram-isca2013, kang-memforum2014}.  We have
shown, in recent works based on rigorous experimental analyses of
modern DRAM chips~\cite{dram-isca2013, samira-sigmetrics14,
  avatar-dsn15, khan-dsn16, memcon-micro17, patel2017reach}, that
determining the minimum retention time of a DRAM cell is getting
significantly more difficult. Thus, determining the correct rate at
which to refresh DRAM cells has become more difficult, as also
indicated by industry~\cite{kang-memforum2014}. This is due to two
major phenomena, both of which get worse (i.e., become more prominent)
with technology scaling. First, Data Pattern Dependence (DPD): the
retention time of a DRAM cell is heavily dependent on the data pattern
stored in itself and in the neighboring
cells~\cite{dram-isca2013}. Second, Variable Retention Time (VRT): the
retention time of some DRAM cells can change drastically over time,
due to a memoryless random process that results in very fast charge
loss via a phenomenon called trap-assisted gate-induced drain
leakage~\cite{yaney1987meta, restle1992dram, dram-isca2013}. These
phenomena greatly complicate the accurate determination of minimum
data retention time of DRAM cells. In fact, VRT, as far as we know, is
very difficult to test for because there seems to be no way of
determining that a cell exhibits VRT until that cell is observed to
exhibit VRT and the time scale of a cell exhibiting VRT does not seem
to be bounded, given the current experimental
data~\cite{dram-isca2013,samira-sigmetrics14,avatar-dsn15,patel2017reach}. As
a result, some retention errors can easily slip into the field because
of the difficulty of the retention time testing.  Therefore, data
retention in DRAM is a vulnerability that can greatly affect both
reliability and security of current and future DRAM generations. We
encourage future work to investigate this area further, from both
reliability and security, {\em as well as} performance and energy
efficiency perspectives.  Various works in this area provide insights
about the retention time properties of modern DRAM devices based on
experimental data~\cite{dram-isca2013, samira-sigmetrics14,
  avatar-dsn15, khan-dsn16, memcon-micro17, softmc, patel2017reach},
develop infrastructures to obtain valuable experimental
data~\cite{softmc}, and provide potential solutions to the DRAM
retention time problem~\cite{raidr, dram-isca2013,
  samira-sigmetrics14, avatar-dsn15, khan-dsn16, memcon-cal16,
  memcon-micro17, darp-hpca2014, patel2017reach}, all of which the
future works can build on.

Note that data retention failures in DRAM are likely to be
investigated heavily to ensure good performance and energy efficiency.
And, in fact they already are being investigated for this purpose
(see, for example,~\cite{raidr, darp-hpca2014, memcon-cal16,
  samira-sigmetrics14, avatar-dsn15, khan-dsn16, memcon-micro17, patel2017reach}).
We believe it is important for such investigations to ensure no new
vulnerabilities (e.g., side channels) open up due to the solutions
developed.


\textbf{NAND Flash Data Retention Issues}

Experimental analysis of modern flash memory devices show that the
dominant source of errors in flash memory are data retention
errors~\cite{cai-date12,cai2017error}. As a flash cell wears out, its
charge retention capability degrades~\cite{cai-date12, cai-hpca15,
  cai2017errors, cai2017error, luo2018improving, luo2018heatwatch,
  justin-flash-sigmetrics15, flash-field-analysis2} and the cell
becomes leakier.  As a result, to maintain the original data stored in
the cell, the cell needs to be refreshed~\cite{cai-iccd12,
  cai-itj2013}. The frequency of refresh increases as wearout of the
cell increases. We have shown that performing refresh in an adaptive
manner greatly improves the lifetime of modern MLC (multi-level cell)
NAND flash memory while causing little energy and performance
overheads~\cite{cai-iccd12, cai-itj2013}. Most high-end SSDs today
employ such adaptive refresh mechanisms.

As flash memory scales to smaller manufacturing technology nodes and
even more bits per cell, data retention becomes a bigger problem. As
such, it is critical to understand the issues with data retention in
flash memory. Our recent work provides detailed experimental analysis
of data retention behavior of planar and 3D MLC NAND flash
memory~\cite{cai-hpca15, cai2017errors, cai2017error,
  luo2018improving, luo2018heatwatch}. We show, among other things,
that there is a wide variation in the leakiness of different flash
cells: some cells leak very fast, some cells leak very slowly. This
variation leads to new opportunities for correctly recovering data
from a flash device that has experienced an uncorrectable error: by
identifying which cells are fast-leaking and which cells are
slow-leaking, one can probabilistically estimate the original values
of the cells before the uncorrectable error occurred. This mechanism,
called {\em Retention Failure Recovery}, leads to significant
reductions in bit error rate in modern MLC NAND flash
memory~\cite{cai-hpca15, cai2017errors, cai2017error} and is thus very
promising. Unfortunately, it also \jk{points to} a potential security
and privacy vulnerability: by analyzing data and cell properties of a
failed device, one can potentially recover the original data. We
believe such vulnerabilities can become more common in the future and
therefore they need to be anticipated, investigated, and understood.

\subsubsection{Other Vulnerabilities in NAND Flash Memory}

We believe other sources of error (e.g., cell-to-cell interference) and
cell-to-cell variation in flash memory can also lead various vulnerabilities.
For example, another type of variation (that is similar to the variation in
cell leakiness that we described above) exists in the vulnerability of flash
memory cells to read disturbance~\cite{cai-dsn15}: some cells are much more
prone to read disturb effects than others. This wide variation among cells
enables one to probabilistically estimate the original values of cells in flash
memory after an uncorrectable error has occurred. Similarly, one can
probabilistically correct the values of cells in a page by knowing the values
of cells in the neighboring page~\cite{cai-sigmetrics14}. These
mechanisms~\cite{cai-dsn15, cai-sigmetrics14} are devised to improve flash
memory reliability and lifetime, but the same phenomena that make them
effective in doing so can also lead to potential vulnerabilities, which we
believe are worthy of investigation to ensure security and privacy of data in
flash memories.

As an example, we have recently shown~\cite{cai-hpca17} that it is
theoretically possible to exploit vulnerabilities in flash memory
programming operations on existing solid-state drives (SSDs) to cause
(malicious) data corruption. This particular vulnerability is caused
by the {\em two-step programming} method employed in dense flash
memory devices, e.g., MLC NAND flash memory. An MLC device partitions
the threshold voltage range of a flash cell into four
distributions. In order to reduce the number of errors introduced
during programming of a cell, flash manufacturers adopt a two-step
programming method, where the least significant bit of the cell is
partially programmed first to some intermediate threshold voltage, and
the most significant bit is programmed later to bring the cell up to
its full threshold voltage.  We find that two-step programming exposes
new vulnerabilities, as both cell-to-cell program interference and
read disturbance can disrupt the intermediate value stored within a
multi-level cell before the second programming step completes. We show
that it is possible to exploit these vulnerabilities on existing
solid-state drives (SSDs) to alter the partially-programmed data,
causing (malicious) data corruption. We experimentally characterize
the extent of these vulnerabilities using contemporary 1X-nm (i.e.,
15-19nm) flash chips~\cite{cai-hpca17}. Building on our experimental
observations, we propose several new mechanisms for MLC NAND flash
that eliminate or mitigate disruptions to intermediate values,
removing or reducing the extent of the vulnerabilities, mitigating
potential exploits, and increasing flash lifetime by
16\%~\cite{cai-hpca17}. We believe investigation of such
vulnerabilities in flash memory will lead to more robust flash memory
devices in terms of both reliability and security, as well as
performance. In fact\jk{,} a recent work from IBM builds on our
work~\cite{cai-hpca17} to devise a security attack at the file system
level~\cite{ibm-fs-attack}.

\subsection{Prevention}
\label{sec:prevention}
\label{sec:future-prevention}

Various reliability problems experienced by scaled memory technologies, if not
carefully anticipated, accounted for, and corrected, may surface as security
problems as well, as in the case of RowHammer.  We believe it is critical to
develop principled methods to understand, anticipate, and prevent such
vulnerabilities. In particular, principled methods are required for three major
steps in the design process.

First, it is critical to understand the potential failure mechanisms and
anticipate them beforehand. To this end, developing solid methodologies for
failure modeling and prediction is critical. To develop such methodologies, it
is essential to have real experimental data from past and present devices. Data
available both at the small scale (i.e., data obtained via controlled testing
of individual devices, as in, e.g.,~\cite{dram-isca2013, aldram, samira-sigmetrics14, kevinchang-sigmetrics16, cai-date12, cai-date13, cai-dsn15, cai-hpca15, yixin-jsac16, cai2017errors, cai2017error, cai-iccd12, cai-itj2013, cai-iccd13, cai-sigmetrics14, cai-hpca17, yucai-thesis, luo2018improving, luo2018heatwatch, patel2019understanding, kim2018solar, patel2017reach, kim2019drange, kim2018dram}) as well as at the large scale (i.e., data
obtained during in-the-field operation of the devices, under
likely-uncontrolled conditions, as in, e.g.,~\cite{justin-memerrors-dsn15,
justin-flash-sigmetrics15}) can enable accurate models for failures, which
could aid many purposes, including the development of better reliability
mechanisms and prediction of problems before they occur.

Second, it is critical to develop principled architectural methods that can
avoid, tolerate, or prevent such failure mechanisms that can lead to
vulnerabilities. For this, we advocate co-architecting of the system and the
memory together, as we described earlier. Designing intelligent, flexible,
configurable, programmable, patch-able memory controllers that can understand and correct existing and
potential failure mechanisms can greatly alleviate the impact of failure
mechanisms on reliability, security, performance, and energy efficiency. A {\em
system-memory co-design} approach can also enable new opportunities, like
performing effective processing near or in the memory device
(e.g.,~\cite{rowclone, vivek-and-or, tesseract, pei, gs-dram, tom-isca16,
impica-iccd16, lazypim, pattnaik-pact16, emc-isca16, 
ghose2018enabling, boroumand2018google, kim2018grim, seshadri2017ambit,
liu2017concurrent, seshadri2017simple, aga2017compute, akin2015data,
asghari2016chameleon, babarinsa2015jafar, lisa, chi2016prime, farmahini2015nda,
gao2015practical, gao2016hrl, gu2016biscuit, guo20143d, cre-micro16,
hassan2015near, hsieh2016accelerating, sramsod, kim2016neurocube,
kim2017toward, lee2015bssync, li2017drisa, pinatubo, loh2013processing,
morad2015gp, nai2017graphpim, pugsley2014ndc, shafiee2016isaac,
sura2015data, zhang2014top, zhu2013accelerating, mutlu2019processing, singh2019napel, mutlu2019enabling, boroumand2019conda}). In addition to designing the
memory device together with the controller, we believe it is important to
investigate mechanisms for good partitioning of duties across the various
levels of transformation in computing, including system software, compilers,
and application software.

Third, it is critical to develop principled methods for electronic
design, automation and testing, which are in harmony with the failure
modeling/prediction and system reliability methods, which we mentioned
in the above two paragraphs. Design, automation and testing methods
need to provide high and predictable coverage of failures and work in
conjunction with architectural and across-stack mechanisms. For
example, enabling effective and low-cost {\em online profiling of
  DRAM}~\cite{dram-isca2013, samira-sigmetrics14, avatar-dsn15,
  khan-dsn16, memcon-cal16, lee2017design, patel2017reach} in a
principled manner requires cooperation of failure modeling mechanisms,
architectural methods, and design, automation and testing methods.

%% file: conclusion.tex
\section{Conclusion}



We provided a retrospective on the RowHammer problem and our original
ISCA 2014 paper~\cite{rowhammer-isca2014} that introduced the problem,
and a survey of many flourishing works that have built on
RowHammer. It is clear that the reliability of memory technologies we
greatly depend on is reducing, as these technologies continue to scale
to ever smaller technology nodes in pursuit of higher densities. These
reliability problems, if not anticipated and corrected, can also open
up serious security vulnerabilities, which can be very difficult to
defend against, if they are discovered in the field. RowHammer is an
example, likely the first one, of a hardware failure mechanism that
causes a practical and widespread system security vulnerability. As
such, its implications on system security research are tremendous and
exciting. We hope the summary, retrospective, and commentary we
provide in this paper on the RowHammer phenomenon are useful for
understanding the RowHammer problem, its context, mitigation
mechanisms, and the large body of work that has built on it in the
past five years.

We believe that the need to prevent such reliability and security
vulnerabilities at heavily-scaled memory technologies opens up new
avenues for principled approaches to 1) understanding, modeling, and
prediction of failures and vulnerabilities, and 2) architectural as
well as design, automation and testing methods for ensuring reliable
and secure operation. We believe the future is very bright for
research in reliable and secure memory systems, and many discoveries
abound in the exciting yet complex intersection of reliability and
security issues in such systems.